%% file: paper-apu-genomics.tex
\documentclass{cbxarticle-ieee}

\usepackage{array}
\usepackage{multirow}
\newcolumntype{P}[1]{>{\raggedright\arraybackslash}p{#1}}
\newcolumntype{C}[1]{>{\centering\arraybackslash}p{#1}}
\newcolumntype{R}[1]{>{\raggedleft\arraybackslash}p{#1}}
\usepackage{breqn}
\usepackage{amsmath}
\usepackage{authblk}

\usepackage{wasysym}

\begin{document}


\pagenumbering{arabic}

\title
{%
  Accelerating Seed Location Filtering in DNA Read Mapping \\
  Using a Commercial Compute-in-SRAM Architecture
}

\papernum{XXX}
\confabbr{CONF XXXX}

\author[1]{Courtney Golden}
\author[2]{Dan Ilan}
\author[1]{Nicholas Cebry}
\author[1]{Christopher Batten}
\affil[1]{School of Electrical and Computer Engineering,
  Cornell University, Ithaca, NY
  \authorcr Email: {\{ckg35, nfc35, cbatten\}@cornell.edu}\vspace{1.5ex}}
\affil[2]{GSI Technologies Inc., Tel Aviv, Israel
\authorcr Email: {dilan@gsitechnology.com}}

\maketitle
\thispagestyle{plain}
\pagestyle{plain}


\input{sec-abstract}

\input{sec-intro}
\input{sec-background}
\input{sec-acceleration}
\input{sec-methodology}
\input{sec-results}
\input{sec-related}
\input{sec-conclusion}


\begin{acknowledgments}
This work was supported in part by NSF PPoSS Award~\#2118709 and NSF SHF Award~\#2008471. The authors acknowledge and thank Santiago Marco, Miquel Moreto, and Max Doblas for their suggestion of using the Myers' bit-parallel algorithm as an algorithm for filtering, and for their help in understanding the biological context of DNA sequence alignment. The authors also thank Caroline Huang and Niansong Zhang for their help in learning how to use the Gemini platform. 
\end{acknowledgments}

\bibliographystyle{cbxabbrv}
\balance
\bibliography{cbatten}

\end{document}

%% file: sec-abstract.tex

\begin{abstract}
DNA sequence alignment is an important workload in computational genomics. Reference-guided DNA assembly involves aligning many read sequences against candidate locations in a long reference genome. To reduce the computational load of this alignment, candidate locations can be pre-filtered using simpler alignment algorithms like edit distance. Prior work has explored accelerating filtering on simulated compute-in-DRAM, due to the massive parallelism of compute-in-memory architectures. In this paper, we present work-in-progress on accelerating filtering using a commercial compute-in-SRAM accelerator. We leverage the recently released Gemini accelerator platform from GSI Technology, which is the first, to our knowledge, commercial-scale compute-in-SRAM system. We accelerate the Myers' bit-parallel edit distance algorithm, producing average speedups of 14.1$\times$ over single-core CPU performance. Individual query/candidate alignments produce speedups of up to 24.1$\times$. These early results suggest this novel architecture is well-suited to accelerating the filtering step of sequence-to-sequence DNA alignment. 
\end{abstract}

%% file: sec-intro.tex

\section{Introduction}
\label{sec-intro}

DNA sequence alignment is an important component of a vast array of biological tasks. In particular, genome assembly involves aligning a collection of read sequences against a long reference sequence. State-of-the-art genome alignment pipelines adopt a seed-and-extend approach, which involves using kmer-matching to first find candidate locations for each read (query) in the reference genome~\cite{li-minimap2-bioinformatics2018, cali-genasm-micro2020}. Then, each query is aligned to its set of candidates using gap-affine approximate string matching, usually a variant of the Smith-Waterman algorithm~\cite{smith-sw-algorithm-jmb1981}. Previous work has suggested that the most computationally expensive step of the pipeline is the gap-affine alignment~\cite{nag-gencache-micro2019}. In order to reduce the time spent in this step, prior work has proposed adding a filtering step to filter the candidates before alignment and reduce the number of candidates per query. A common approach to filtering is computing a simpler edit-distance score for each query/candidate pair that approximates the similarity between the sequences. Candidates whose scores are above a certain threshold are then filtered out~\cite{kim-grimfilter-2018, alser-shouji-bioinformatics2019, alser-gatekeeper-bioinformatics2017}. Thus, adding filtering and accelerating it in hardware can effectively reduce the number of alignments performed later in the pipeline and improve overall end-to-end genome assembly execution time. 

Recent advances in compute-in-memory technologies make it a promising architecture on which to accelerate such computational genomics algorithms. Compute-in-memory (CIM) architectures seek to reduce computation time spent on data movement by closing the traditional gap between processors and storage elements, which is highly beneficial for data-intensive workloads like DNA alignment. Prior work by Nag et al.~proposed GenCache, a tightly-coupled compute-in-SRAM accelerator with hardware extensions for filtering operations~\cite{nag-gencache-micro2019}. GenCache improves execution time of end-to-end alignment (including filtering and other steps) by 5.26$\times$ over an identical accelerator without the added in-cache capabilities. Another work by Kim et al.~proposed GRIM-Filter~\cite{kim-grimfilter-2018}, which also uses CIM to accelerate the filtering step of DNA sequence alignment pipelines. GRIM-Filter explores using loosely-coupled, 3D-stacked DRAM architectures for short-read assembly with custom hardware for filtering. Their system improves overall end-to-end genome assembly time by up to 3.65$\times$. 

Inspired by these promising, simulation-based studies, we are exploring the potential for using a general-purpose, commercial compute-in-SRAM architecture to accelerate the filtering step of short-read DNA assembly. Relative to DRAM, the physical scaling of SRAM has more closely tracked transistor and logic scaling, so compute-in-SRAM enables monolithic solutions with lower area- and energy-overheads. The Gemini accelerator platform from GSI Technologies is a recently-released commercial system~\cite{gwennap-gsi-linley2020}. To our knowledge, it is the only commercial-scale compute-in-SRAM chip. It offers direct computing capabilities in a 1.5MB SRAM bank, supported by three other layers of memory hierarchy, and communicates with an x86-64 host processor over PCIe. It's highly efficient search and update operations across large arrays of data give it the name Associative Processing Unit, or APU. We thus present our on-going work suggesting this novel architecture is well-suited to accelerating the filtering step of sequence-to-sequence genome alignment. 

In this paper, we detail the architecture and programming model of the Gemini APU accelerator platform and then use it to accelerate the Myers' bit-parallel edit-distance algorithm, as originally proposed by G. Myers in~\cite{myers-algorithm-1999}. The Myers' algorithm provides an efficient approach to edit-distance calculation. Using simple bit-operations, the algorithm calculates a score for each query/candidate pair indicating the number of base-pair edits needed for the best alignment of that pair. Candidates with scores above a certain threshold can be filtered out prior to the subsequent expensive gap-affine alignment step. Like GRIM-Filter, we focus on short-read DNA assembly, with query sequences that are hundreds of base pairs in length. We evaluate the performance of the APU relative to a Intel(R) Xeon(R) Gold 6230R CPU using a dataset consisting of 500 simulated queries of length 300 base pairs (bp) and corresponding candidate sequences generated from a standard human genome reference sequence. The APU provides speedups of up to 14.1$\times$ for the entire dataset, with speedups of up to 24.1$\times$ for individual queries with large numbers of candidates. 

Our key contributions include: (1) the first publicly-available, microcode-level description of the architecture and programming model of the Gemini APU accelerator platform, and (2) early work demonstrating the ability of the APU to accelerate the filtering step of the DNA short-read assembly pipeline using the Myers' bit-parallel edit-distance algorithm. 

%% file: sec-background.tex

\section{Gemini Accelerator Architecture}
\label{sec-background}

\begin{figure*}
    \centering
    \includegraphics[width=\textwidth]{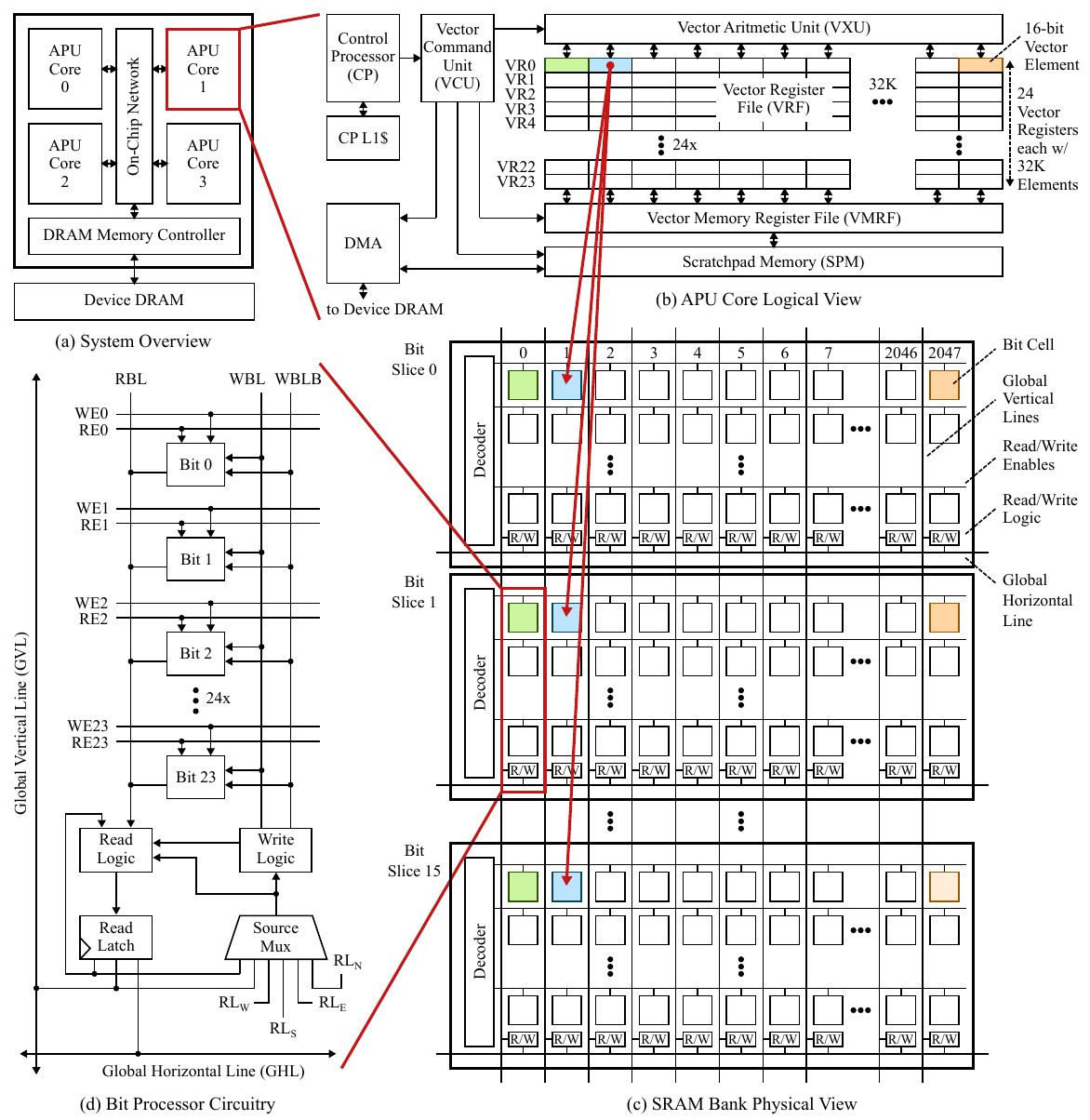}
    
    \caption{APU Architecture -- (a) System Overview, (b) APU Core Logical View, (c) Bank Physical View, (d) Bit Processor Circuitry. CP = control processor, VCU = vector command unit, VXU = vector execution unit, VRF = vector register file, VMRF = vector memory register file, SPM = scratchpad memory, GVL = global vertical latch, R/W = read/write logic, RBL = read bitline, WBL = write bitline, WBLB = write bitline bar, REx = read-enable for bit x, WEx = write-enable for bit x, RL\textsubscript{N} = north read latch. Note: exact bit-slice organization is not published by GSI.}
    \label{fig:archdiag}
\end{figure*}

In order to provide insights into the features of the Gemini APU platform that make it well-suited to accelerating the Myer's bit-parallel algorithm and other potential genomics computations, in the following section we present a simplified view of the APU's architecture, microarchitecture, and programming model (see Figure \ref{fig:archdiag}). There is currently no detailed public description providing a microcode-level view of the accelerator platform. Here, we detail the main features that affect microcode programming, while abstracting away some additional functionalities not relevant to the results in this paper. See~\cite{gwennap-gsi-linley2020} for more high-level information about the APU. In order to provide the best intuitive description of the Gemini platform's microarchitecture, we have adopted slightly different terminology than GSI Technology uses. See Table \ref{tab:terms} for a conversion between the terms we use here and the ones found in GSI's published materials.

\subsection{System Overview}

The Gemini platform consists of a host CPU and an APU chip. The host stages data by copying values to a 16GB shared DDR4 DRAM located on the device, and then launches a kernel to the device. The APU chip contains four APU cores connected by an on-chip network, and a control processor inside each accelerator core runs the kernel issued by the host. 

\subsection{Accelerator Core Logical View}

Logically, each accelerator core can be viewed as a vector engine containing a control processor (CP), a vector register file (VRF), vector execution logic (VXU), and a selection of other memory and control blocks (Figure \ref{fig:archdiag}b). Within a core, instruction distribution starts in the 32-bit control processor, which executes scalar code and issues vector instructions to the vector command unit (VCU). The VCU decodes vector instructions, like vector-vector addition, into microcode operations that directly control the hardware at a cycle-level granularity. The VCU can issue a single instruction per cycle that operates on all 16 bits of each element, or it can generate up to four instructions per cycle that can operate on subsets of the 16 bits. 

Data flows from the host processor in a series of transfers through a memory hierarchy. Data in the shared device DRAM can be accessed by the device using a special pointer-like mechanism called a memory handle. The accelerator cores use direct memory access (DMA) to transfer data from the device DRAM to their 64KB scratchpad memories (SPM), which are local to each core. Sequences of microcode issued by the VCU then transfer data closer to the execution units inside a 3MB vector memory register file (VMRF), which serves as a set of 48 "background" registers. Ultimately, the main unit of local storage (the compute-in-memory block) is a 1.5MB vector register file (VRF). Data is most naturally organized as 24 vector registers (VRs) that each contain 32,768 16-bit elements. 15 of these VRs are exposed to the application C code, while the remaining are reserved as temporaries for microcode functions. 

To simplify communication and data movement, the VRF is divided into 16 banks, each of which contain 2048 of the 32K elements in each vector register. Elements of a vector register are striped across the banks. SRAM cells within each bank share read- and write-enable signals, and adjacent SRAM cells in any direction can easily transfer data.

\begin{table}
  \centering
  \cbxsetfontsize{9pt}
  \caption{Microarchitecture Terminology}
  \label{tab:terms}

  \begin{tabular}{ll}
  \toprule
       \BF{Term Used in This Work}        & \BF{GSI Term} \\\midrule
       Vector Register File (VRF)         & MMB           \\
       Vector Memory Register File (VMRF) & L1            \\
       Scratchpad Memory (SPM)            & L2            \\
       Device DRAM                        & L4            \\
       Control Processor (CP)             & ARC           \\
       SRAM Bit-Slice                     & Section       \\
       Global Vertical Line (GVL)         & GL            \\
       Global Horizontal Line (GHL)       & RSP           \\
  \bottomrule
  \end{tabular}
\end{table}

\subsection{Bank Physical View}

Figure \ref{fig:archdiag}c shows the actual microarchitecture of one bank of the VRF. There are 2048 columns, each of which stores the 16 bits of an element vertically. Data is stored in a bit-sliced fashion, where corresponding bits of each VR are stored together as one bit-slice. For example, in column zero, bit-slice $i$ contains bit $i$ of element zero of all 24 vector registers, bit-slice $i+1$ contains bit $i+1$ of element zero from all 24 vector registers, etc. Each bit processor (one column of a bit-slice) thus contains 24 bits, along with some associated digital logic to perform computations on those bits. The green elements in Figure \ref{fig:archdiag} indicate how the 16 bits of the first element of the first VR are stored in a bit-sliced manner across the bit-slices. The blue and orange elements show how the bits of two other elements of the first VR map to the 16 bit-slices, where various elements from across the VR are stored in bank 0 due to striping elements across banks. Each bit-slice contains 2048 bit processors (BP) that have their own bitline logic but share read- and write-enables. 

To facilitate global communication, there is a global vertical latch (GVL) that connects all bit-slices in a column. This latch enables data broadcasts, logical ANDs of multiple elements, and bitwise shifts with wraparound. The architecture also includes a global horizontal line (GHL) for each bit-slice. While its actual microarchitectural implementation is more complex, the GHL functionality ORs together the read latches of all the elements in each bit-slice. 

We have described here the most intuitive way to store data: the 16 bit-processors in a column are used to store a single 16-bit element of each vector register in a bit-sliced fashion, and the 32K columns store 32K elements. However, data can be stored in many other formats. For example, data of smaller bitwidths could be stored while increasing the vector length, by storing multiple smaller elements in a single column. Alternatively, wider elements such as 32-bit values could be stored using two columns per vector element with a 16K vector length. To increase throughput for bit-serial operations, each bit-slice could be assigned a different vector element, with the bits of the element loaded from the VMRF in subsequent cycles. In this work, we focus only on the canonical data organization. 

\subsection{Bit Processor}
\label{subsec-bitprocessor}

Inside each bit processor (BP), bits are stored in custom 12T SRAM cells. A small collection of logic gates is found at the bottom of each bit processor ("R/W Logic" in Figure \ref{fig:archdiag}(d), functionally equivalent to the VXU in the micro-programmer view in Figure \ref{fig:archdiag}(b) above). As demonstrated in Figure \ref{fig:archdiag}(d), to operate on any given bit, data values are read from memory cells using a single common read bit-line (RBL), and, after optionally performing a simple logical operation in the read logic, are stored in a read latch (RL). The read logic can perform AND, OR, and XOR on two or more operands, including the VRF, RL, and GVL. To perform arithmetic operations, the hardware enables data transfer between adjacent bit processors by allowing each bit processor to read from its own read latch along with the read latches of the bit processors to its north, south, east, and west (i.e., RL\textsubscript{N}, RL\textsubscript{S}, RL\textsubscript{E}, and RL\textsubscript{W}). In addition, multiple rows can be read from the memory array in a single cycle; the logical AND of the values appears on the RBL. Write operations choose from the operands in the same source mux and modify the VRF using the write bit-line (WBL) and its negation (WBLB). 

By default, the same operations are performed in all 16 bit-slices and all 2048 columns simultaneously. However, different operations can be done on different bit-slices at once using a bit-mask, a 16-bit value indicating which bit-slices to operate on. For instance, one microcode instruction could read a subset of the bits from one VR, and then a second microcode instruction could read the remaining bits from a the remaining bits from a different VR. 

\subsection{Microcode}

\begin{table}
  {\footnotesize
  \centering

  \caption{Microcode Semantics}
  \label{tab:microcode}

  \newcommand{\ii}[1]{{\normalfont\IT{#1}}}

  \setlength{\tabcolsep}{3pt}
  \begin{tabular}{>{\ttfamily}lp{5.6cm}}
  \toprule
      \multicolumn{2}{l}{\BF{Microarchitectural State}} \\\midrule
      RL          & read latch \\
      GVL         & global vertical latch \\
      GHL         & global horizontal latch \\
      VRF[\ii{i}] & vector register source \IT{i} in VRF \\[0.05in] \toprule

      {\normalfont\BF{Operations on State}} \\\midrule
      RL = VRF[\ii{vrs0}]                      & read vector register from VRF \\
      RL = VRF[\ii{vrs0},\ii{vrs1}]            & read two vector registers, bitwise AND the values \\
      RL = \IT{L}                              & read value from a source latch \\
      RL = VRF[\ii{vrs0}] \ii{op L}      & operate on new values from VRF and a latch \\
      RL \ii{op}= VRF[\ii{vrs0}]               & operate on current RL and new value from VRF	\\
      RL \ii{op}= \ii{L}                       & operate on current RL and new value from latch	\\
      RL \ii{op}= VRF[\ii{vrs0}] \ii{op L}     & operate on current RL, value from VRF, and latch \\
      VRF[\ii{vrs0}] = \ii{L}                  & write to VRF from source latch \\[0.05in] \toprule

      {\normalfont\BF{Bit Masking}} \\\midrule
      \ii{bm}:\ii{ stmt}             & 16-bit mask (\IT{bm}) activates bit-slices \\
      (\ii{bm} <{}< \ii{imm}):\ii{ stmt}  & \IT{bm} can be bitwise shifted by immediate (\IT{imm})	\\
  \bottomrule
  \end{tabular}\par}

  \vspace{0.05in}
  \footnotesize

  \IT{L} is latch specifier (i.e.,
  \TT{RL}, \TT{GVL}, \TT{GHL}, \TT{RL\tsub{N}}, \TT{RL\tsub{S}}, \TT{RL\tsub{E}},
  \TT{RL\tsub{W}}); also possible to use complement of latch (i.e., \TT{\ttilde{}RL}, \TT{\ttilde{}GVL},
  \TT{\ttilde{}GHL}, \TT{\ttilde{}RL\tsub{N}}, \TT{\ttilde{}RL\tsub{S}}, \TT{\ttilde{}RL\tsub{E}}, \TT{\ttilde{}RL\tsub{W}}).

\end{table}

Programmer-visible function calls are decoded into microcode instructions by the VCU. These microinstructions control read and write logic, global structures (GVL and GHL), and data transfer between memory layers (VRF, VMRF, and SPM). Table \ref{tab:microcode} gives an overview of the syntax and semantics of the subset of microcode operations that we use in this paper. 

An APU core has three main types of microarchitectural state: read latches, global latches, and the VRF. Read latches allow operations on values currently stored inside this bit processor's circuitry, or values stored in its four neighboring bit processors. As introduced above, the GVL connects the 16 read latches in a column while the GHL ORs together all 32K read latches in each bit-slice. Bit processors also contain circuitry to negate values as they are read or written. Read operations store a new value in the read latch by performing AND, OR, or XOR operations on either existing values in the latch or new values selected by a source mux. Write operations store data values back to the VRF. 

Each of these operations can be expressed as a line of microcode with the inclusion of a bit-mask. Additionally, although we do not utilize this functionality in this paper, up to four micro-operations can be combined into a single VLIW instruction issued by the VCU. It is the responsibility of the programmer to avoid structural hazards when writing microcode sequences. 

%% file: sec-acceleration.tex

\section{Acceleration of Myers' Bit-Parallel Algorithm}
\label{sec-acceleration}

Reference-guided genome assembly involves a series of computational steps. First, candidate locations in the reference genome are identified for each query sequence. Optionally, these candidate locations can be chained together into longer segments or filtered to reduce the number of candidates per query. Then, each query/candidate pair is aligned using an approximate string matching algorithm, and the final assembly is put together. In this work, we focus on the filtering step. Effective and highly-performant filtering can reduce the time spent in the later alignment stage. 

\subsection{Short-Read Alignment Filtering}


In the filtering step of short-read DNA alignment, the goal is to compare each read (short DNA sequence) in a dataset against many corresponding candidate locations using one-to-many alignment. Candidates are generated using a seed-and-extend technique. For a given read, this method involves first pre-computing a hash table that stores the locations of kmers (small subsequences) in the reference genome, and then indexing into the hash table to find exact match locations for all kmers located in the read. An extended region around each of these exact matches is gathered, creating a set of queries and for each one, a set of potential candidate locations in the reference. The output of filtering is a score for each query/candidate pair equal to the minimum number of edits needed to make that query and candidate align exactly. For a given query, all candidates are of equal length, but query and candidate lengths can vary for different reads from the same genome. 

\subsection{Myers' Bit-Parallel Algorithm}


Myers' bit-parallel algorithm computes the edit distance between two DNA sequences, measuring the minimum number of base-pair edits to the sequences that would make the two strings match exactly at their best relative alignment location~\cite{myers-algorithm-1999}. Edits can consist of insertions, deletions, or substitutions of base pairs~\cite{kim-grimfilter-2018}. 

The traditional approach to computing edit distance requires making a large matrix C of size \((m + 1)\, \times\, (n + 1)\), where $m$ is the number of base pairs in the query and $n$ is the number of base pairs in each candidate. Each element \((i,j)\) is computed as follows:
\begin{align*}
    C\,[i, \,j] = min \{ \,\,C\,[i-1,\, j-1] \,+ \,!\,(r_{i} == s_{j}), \\
                   C\,[i-1, \,j]\, +\, 1, \,
                   C\,[i, \,j-1] \,+ \,1 \}
\end{align*}
where $r_{i}$ is the $i$th base pair of the query and $s_{j}$ is the $j$th base pair of the candidate. $C\,[i, \,j]$ depends on the elements above, to the left, and diagonally to the upper left of the desired element \((i,j)\). Each column (corresponding to a base pair of the candidate) produces a score along the bottom row of the matrix. When using edit distance for filtering in DNA alignment, the output is a single scalar score for each query/candidate pair, equal to the minimum value in this last row of scores. 

\begin{figure*}

    \begin{minipage}[t]{0.17\textwidth}
        \footnotesize
        \begin{verbatim}
(a)
0  let m = length(query)
1  let n = length(candidate)
2
3  for q in [0...num_queries-1]:
4    precompute peq
5  
6    for c in [0...num_cand-1]:
7      Pv = 1^m
8      Mv = 0^m
9      score, min = m
10         
11     for j in [0...n-1]:
12       eq = peq[seed(j), i]
13       Xv = eq | Mv
14       Xh = ((eq & Pv) 
               + Pv ^ Pv) | eq
15       Ph = Mv | ~(Xh | Pv)
16       Mh = Pv & Xh
17 
18       if Ph MSB = 1, score += 1
19       if Mh MSB = 1, score -= 1
20       if score < min, min = score
21                 
22       shift Ph
23       save old MSB of Ph
24 
25       shift Mh
26       save old MSB of Mh
27 
28       Pv = Mh | ~(Xv | Ph)
29       Mv = Ph & Xv

\end{verbatim}
\end{minipage}
    \hfill
    \begin{minipage}[t]{0.18\textwidth}
        \footnotesize
        \begin{verbatim}
(b) 
APL_FRAG vor_vv(vrd, vrs0, vrs1): 
   0xFFFF: RL = VRF[vrs0];
   0xFFFF: RL |= VRF[vrs1];
   0xFFFF: VRF[vrd] = RL;
\end{verbatim}

        \begin{verbatim}
(c) 
APL_FRAG vmseq(vrd_m, vrs, rs):
  0xFFFF: RL = VRF[vrs];
  ~rs: RL = ~RL;
  0xFFFF: GVL = RL;
  vrd_m: VRF[MASK_REG] = GVL;
\end{verbatim}
\begin{verbatim}
(d)
APL_FRAG vmv_vx(vrd, in_value):
  0xFFFF: RL = 0;
  in_value: RL = 1;
  0xFFFF: VRF[dst] = RL;

(e)
APL_FRAG save_last_bit(vrd, b16, bit):
  b16: GVL = RL;
  bit: VRF[dst] = GVL;

(f)
APL_FRAG lsl_with_cin(vrs, shift_in):
  0xFFFF: RL = VRF[src];
  0xFFFF: VRF[src] = NRL;
  0x0001: RL = VRF[shift_in];
  0x0001: VRF[src] = RL;

\end{verbatim}
    \end{minipage}
    \hfill
    \begin{minipage}[t]{0.36\textwidth}
        \footnotesize
        \begin{verbatim}
(g) 
APL_FRAG vadd(vdst, vsrc0, vsrc1):

  // ---- bit 0 ----
  
  // vdst = vsrc0 XOR vsrc1
  0x0001:  RL = VRF[vsrc0];
  0x0001:  RL ^= VRF[vsrc1];
  0x0001:  VRF[vdst] = RL;
  
  // cout = vsrc0 AND vsrc1
  0x0001:  RL = VRF[vsrc0, vsrc1];

  // ---- bit 1 ----
  
  // vdst = a ^ b ^ cin
  (0x0001<<1):  RL = VRF[vsrc0];
  (0x0001<<1):  RL ^= VRF[vsrc1];
  (0x0001<<1):  RL ^= RL_N;
  (0x0001<<1):  VRF[vdst] = RL;
  
  // cout = a*b + b*cin + a*cin
  (0x0001<<1):  RL = VRF[vsrc0, vsrc1];
  (0x0001<<1):  VRF[temp_0] = RL;
  (0x0001<<1):  RL = VRF[vsrc1];
  (0x0001<<1):  RL &= RL_N;
  (0x0001<<1):  VRF[temp_1] = RL;
  (0x0001<<1):  RL = VRF[vsrc0];
  (0x0001<<1):  RL &= RL_N;
  (0x0001<<1):  RL |= VRF[temp_0];
  (0x0001<<1):  RL |= VRF[temp_1];
  ...                
\end{verbatim}
    \end{minipage}
    \begin{minipage}[t]{\textwidth}
        \cbxsetfontsize{9pt}
        \vspace{-.7in}
        \caption{Pseudocode and Microcode for Myers' Algorithm. (a) Pseudocode for alignment of a single read against a single seed from~\cite{myers-algorithm-1999}. (b-g) Microcode fragments for Myers' algorithm implementation: (b) bitwise OR, used in lines 13-15 and 28 of the pseudocode; (c) vector set-equals, used in a multi-line equivalent to line 12 of the pseudocode on Gemini; (d) sets all elements to scalar value, used for data initialization on lines 7-9 and 12; (f) saves the last bit (bit 16) that is currently stored in RL, used directly after the functions appearing in lines 23 and 26 of the pseudocode; (f) left bitwise shift with a carry-in, used in lines 22 and 25; (g) ripple-carry elementwise addition for line 14, although the actual implementation of addition used in our algorithm uses a more sophisticated carry-select approach not shown here. vrd = destination vector register, vrs = source vector register, vrd\_m = a one-hot encoding for which of the 16 bits of each element of MASK\_REG corresponds to the desired mask. rs, in\_value, and shift\_in are all 16-bit integer operands. b16 is a 16-bit value with only the 16th bit set, and bit is a 16-bit mask for a desired bit. }
        \label{fig:microcode}
    \end{minipage}.
\end{figure*}

To reduce the space and time complexity of such an algorithm, Myers adopts a bit-vector approach. Instead of storing the score for every element \((i,j)\) in the matrix, Myer's bit-parallel algorithm stores the horizontal and vertical deltas between adjacent elements. These deltas only have values in the set \{$+1$, 0, $-1$\} and thus can each be represented using two bits. The algorithm iterates over the base pairs of the candidate, as if proceeding column-wise through the score matrix. Vectors \(Pv\) and \(Mv\) are defined as the positive and negative components of the vertical deltas in a given column: at each element \(i\), \(Pv(i) = 1\) and  \(Mv(i) = 0\) if the vertical delta at \((i,j)\) is $+1$ (i.e., \(C[i,j]\) is one more than the element above it), \(Pv(i) = 0\) and  \(Mv(i) = 1\) if the vertical delta at \((i,j)\) is $-1$, and \(Pv(i) = Mv(i) = 0\) if the vertical delta at \((i,j)\) is 0. Similarly, \(Ph\) and \(Mh\) hold the positive and negative components of the horizontal deltas. Then, all calculations of the score deltas can be performed via simple bit-wise operations on these bit-vectors. The use of such simple bitwise operations and the high degree of parallelism in the Myer's algorithm makes it well-suited to acceleration on the APU hardware. The pseudocode for Myers' algorithm can be seen in Figure \ref{fig:microcode}(a). 

Each element of each vector in the pseudocode would ideally have a number of bits equal to the length of an entire column of the score matrix, i.e., the number of base pairs in the query. For short-read assembly, each query is hundreds of base pairs long. However, in a CPU implementation of Myers' algorithm running on a 64-bit processor, we can only pack up to 64 bits of each vector into a single word, and on the APU, as explained below, we operate on 16-bit values. Thus, to support query lengths of greater than \(w\) base pairs (where \(w\) is a CPU's word width or 16 for the APU), we iterate over \(w\)-bit chunks of the query base pairs. In order to correctly account for vertical deltas that propagate down a column of the score matrix, we must save values between iterations of this inner loop. For a particular iteration of the outer loop, corresponding to some base pair at index \(j\) of the candidate, three bits of information per candidate must be saved between iterations: the carry-out  bit of addition in line 11 of the pseudocode in Figure \ref{fig:microcode}(a), the most-significant bit of Ph shifted out in line 23, and the most-significant bit of Mh shifted out in line 26. In our CPU implementation of Myers' algorithm, we use a word width of 64 bits. For each query in the dataset, we loop over all the associated candidates, computing all columns in series for each candidate. 

\subsection{Mapping the Myers' Algorithm to the APU}

\begin{figure}
    \centering
    \includegraphics[width=\columnwidth]{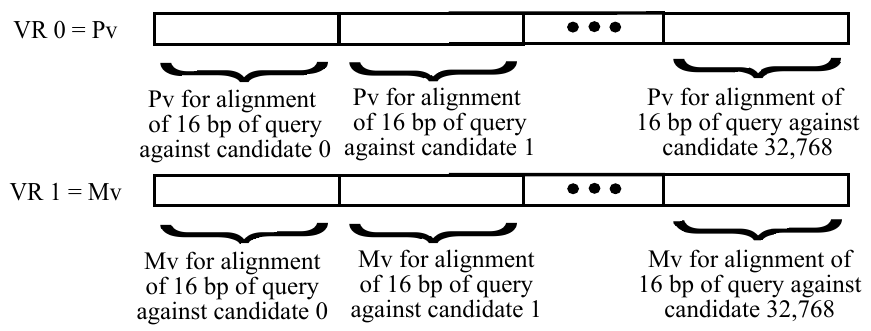}
    \caption{Data Layout of Myers' Algorithm on the APU}
    \label{fig:layout}
\end{figure}


On the APU, we adapt this algorithm to suit the hardware's massive parallelism. Since each of the 32K columns of the APU's SRAM can hold 16 bits, we can perform 512k bit operations in parallel on every cycle. In order to exploit this, we modify the data layout of our CPU-optimized version of Myers' algorithm. For each query, we compute the scores for all candidates in parallel. Specifically, each candidate \(i\) is assigned to column \(i\) in the SRAM (element \(i\) of each of its vector registers). The vector registers hold vectors \(Pv\), \(Mv\), etc., where each element of a VR holds data for 16 base pairs of the read (forming a 16-bit element). The data layout can be seen in Figure \ref{fig:layout}. As a note, this means that the candidate matrix must be generated in a transposed format relative to the input candidate matrix of the CPU version of the algorithm. 


The vector register file of the APU is limited in size, with 15 user-facing vector registers if each holds 32K 16-bit data elements. All 15 of these vector registers are used in each iteration of the innermost loop of the APU implementation of the algorithm, which loops over 16-bit chunks of the query base pairs. However, the vertical deltas computed for a set of 16 query base pairs and one particular candidate base pair must be used in the calculation for the next candidate base pair and that same set of 16 query base pairs. This means the data must be stored temporarily in the VMRF, which provides space to hold up to 48 vector registers' worth of data. Data is moved in and out of these VMR's using microcode. Moving data between just the VMRF and VRF is well-suited for performing short-read alignments, as we do here. Future work can explore long-read alignments, and would have to carefully manage data movement between the VMRF and DRAM. 


The accelerator requires the query and candidate sequences at the start of the algorithm, and produces a result matrix at the end. The time spent copying data from the host to the device is dominated by moving the candidate sequences, so we adopt a bit-packed approach to storing the candidate sequence data. In particular, since each base pair of a candidate sequence can only hold four possible values (corresponding to A, C, T, and G), it can be represented with just two bits. We thus pack eight base pairs into each 16-bit element of candidate sequence data moved between the host and device. Doing so also reduces the time spent in the computational kernel on the accelerator itself, as it requires eight times fewer loads from the shared DDR4 DRAM in favor of simple bitwise shifts (an operation performed highly efficiently by the APU). 


In addition, to further reduce the time spent in computation, we leverage the ability of the APU to quickly initialize data inside its SRAM. By using a 16-bit value as a bit-mask, the device can write this value to all 32K 16-bit elements simultaneously in just three cycles. As a result, the APU can perform massively parallel in-memory initialization, which can save significant amounts of time over a CPU that must instead iterate through a large array.

\subsection{Microcode for the Myers' Algorithm}

\begin{table}
  {\cbxsetfontsize{9pt}
  \centering

  \caption{Microcode Operations Used In Myers'}
  \label{tab:microcode-list}

  \setlength{\tabcolsep}{3pt}
  \begin{tabular}{p{3.8cm}p{4.7cm}}
  \toprule
      \BF{Bitwise Operations} & \BF{Elementwise Operations} \\ \midrule
      bitwise OR  & addition \\
      bitwise AND & less-than comparison \\
      bitwise XOR &  set to scalar value \\
      bitwise NOT & masked vector copy \\ 
      bitwise left-shift w/ carry-in & broadcast bit to entire element \\ 
      extract a desired bit \\
      save bit 16 currently on RL \\
      \midrule
      \BF{Memory Operations} & \BF{Cross-Element Operations} \\ \midrule
      load from VMRF & vector set-equals (search) \\
      store to VMRF  & vector set-equals (search) \\
      load from DRAM \\
      store to DRAM \\
  \bottomrule
  \end{tabular}\par}
\end{table}


In order to implement this approach on the APU, we create C-function wrappers around microcode fragments that execute vector operations corresponding to each line of the pseudocode in Figure \ref{fig:microcode}(a). This includes (1) bitwise vector operations like bitwise OR and bitwise AND, (2) element-wise vector operations, (3) memory operations, and (4) cross-element operations. In addition to well-known arithmetic and Boolean operations, we developed several custom operations, including one that saves the last value computed on the RL of a previous operation or extracts a desired bit from every element of the vector register in parallel. A complete list of all the microcode operations developed or used in this Myers' algorithm implementation can be found in Table \ref{tab:microcode-list}. 


The microcode implementation of bitwise OR is shown in Figure \ref{fig:microcode}(b). Each microcode instruction is executed on every bit-processor across all 32K elements of the vector register file in parallel. As described in Section \ref{sec-background}, corresponding bits of each element are read out from the VRF and operated on by the R/W logic containing an OR gate at the bottom of the bit processor. We use an \TT{|=} operation since the RL is only a 1-bit storage element. The resulting bit is written to the destination VR. In addition to the three microcode instructions shown in Figure \ref{fig:microcode}(b), three additional instructions are needed before that microcode fragment is issued to set registers in the control processor indicating source operands. 


In addition to bitwise operations, the conventional Myers' algorithm implementation also requires elementwise addition between vectors. The microcode shown in Figure \ref{fig:microcode}(g) shows one way to do addition, using a naive bit-serial ripple-carry algorithm, although the actual implementation used in our Myers' algorithm uses a more sophisticated carry-select approach. However, we explain the basic ripple-carry approach because it demonstrates fundamental properties of this architecture. In the bit-serial version shown, for inputs $a$ and $b$, carry-in $c_{in}$, result $r$, and carry-out $c_{out}$, we express the result and carry-out of each bit using the Boolean logic functions \(r_i = XOR(a_i, b_i, {c_{in}}_i)\) and \({c_{out}}_i = a_i * b_i + b_i * {c_{in}}_i + a_i * {c_{in}}_i\). We then implement these expressions using bitlines and peripheral logic circuitry. For each bit, the result bit is first computed and stored, and then a carry-out is computed and propagated to the next bit-slice. The code for bit 1 in Figure \ref{fig:microcode}(g) is replicated, shifting the bit-mask by an additional bit each time, for the remaining 14 bits. This basic implementation makes the simplifying assumption that the source operands are distinct, but identical operands could be easily supported. Our experiments utilize a more sophisticated carry-select approach provided by GSI Technology. 


Accelerating the Myers' algorithm on the APU allows us to take advantage of element-wise search, which is extremely efficient on its associative-style SRAM array. As described in the microcode section below, a 16-bit scalar value can be compared against all elements of a 32K-element vector register in just four microcode instructions, corresponding to roughly four cycles on the device's control processor. This is useful in creating the \(eq\) vector, whose elements hold one-hot encodings of where each candidate sequence's current base pair (A, C, G, or T) appears in the current 16-bit chunk of the query sequence. A precomputed array of where each base pair appears in the query sequence is created on the host processor. Then, for each index \(j\) of the main loop of the algorithm, the device must load different chunks of this data into elements of the \(eq\) vector register corresponding to the different base pairs located at the index \(j\) of each candidate. Such an operation can be done efficiently by searching a vector register containing the \(j\)th base pairs of each candidate and using vector mask operations to load data from \(peq\) into a single \(eq\) vector. 

Figure \ref{fig:microcode}(c) shows the code for this element-wise search (a set-equals operation), which compares a scalar operand to each element of a vector in parallel and marks the elements that match. After reading in all 32K elements in parallel, we selectively negate bits for which the desired scalar operand has a zero. Using the GVL as a reduction element, we AND together the resulting data, which yields a one if the 16-bit element matches the scalar operand and a zero if it differs. 


In order to store the results of a comparison instruction like set-equals, we provide support for representing vector masks. To reduce space overhead, we follow GSI Technology's convention of packing up to twelve masks into a single designated vector register, denoted MASK\_REG. The $i$th mask is stored as 32K bits, where the $i$th bit of each 16-bit element of MASK\_REG corresponds to this mask. Eight masks are exposed to the user for general-purpose storage, while four are used for temporaries by the vector functions. The remaining four bits of each element of MASK\_REG are reserved for use as arithmetic flags.  


The implementation of Myers' described here has many possibilities for future optimizations. For example, as described in~\cite{myers-algorithm-1999}, the addition operator used in line 11 of the Myers' pseudocode is incorporated only to allow for the propagation of carry bits across bits of a single data value in a traditional processor. Since the APU architecture provides efficient low-level support for bitwise shifts and carries, the calculation of \(Xh\) could be reduced to a much simpler set of bit operations that could be done efficiently on the accelerator. Future work can incorporate such optimizations.

%% file: sec-methodology.tex

\section{Evaluation Methodology}
\label{sec-methodology}


To accurately evaluate the performance of the Myers' algorithm on the APU and a CPU baseline, we generate a set of short queries of length 300 base pairs (bp) using a state-of-the-art simulator. Like~\cite{fujiki-genax-isca2018, nag-gencache-micro2019, kim-grimfilter-2018}, we generate test data from the standardized GRCh38 release of the human genome. We used the Mason simulator~\cite{holtgrewe-mason-2010} to generate 500 simulated queries of length 300bp from the reference genome. We configured Mason to simulate queries based on the Illumina sequencing technology characteristics.

Candidate alignments for these queries were then generated with a program written using the SeqAn library~\cite{reinert-seqan-2017}. First, we created a hash-table based index of the reference genome using 10-mer minimizers (short sequences of 10 base pairs that occur in the reference genome). We eliminated any 10-mers that occurred more than 100,000 times in the reference from the index, as these matches occur too frequently to provide useful information in determining a best alignment with a query sequence. This eliminated less than 0.1\% of minimizers in the table, but reduced the number of stored locations by 3.7\%.

\begin{figure}
    \centering
    \includegraphics[width=\columnwidth]{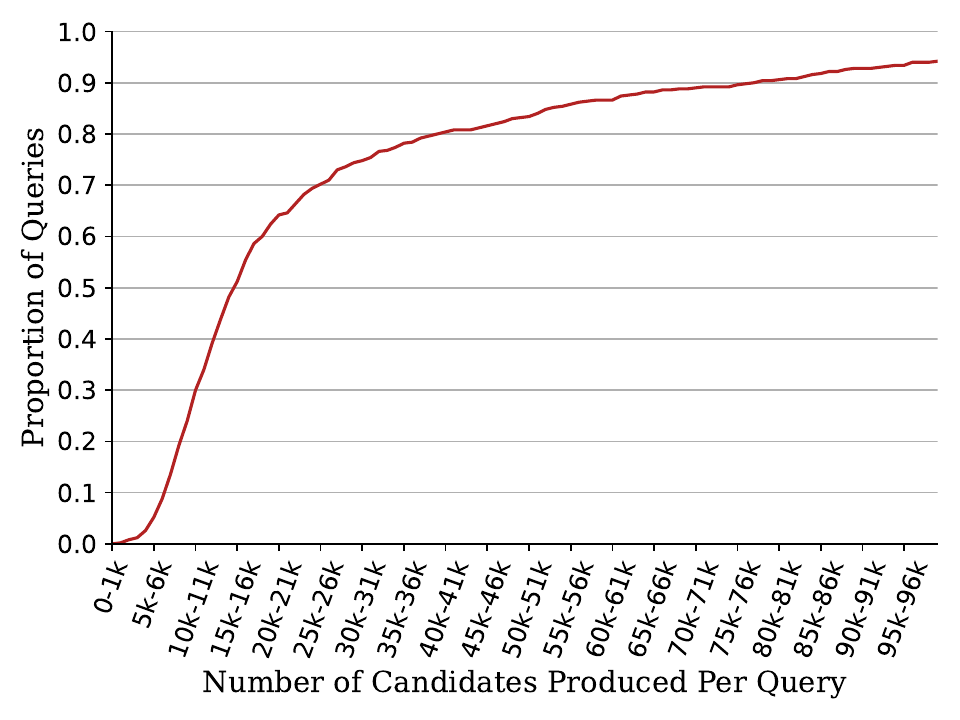}
    \caption{Cumulative Distribution of Candidates Per Query -- Distribution of the number of candidates produced in each length-range (measured in base pairs) for the 300-base pair reads produced by Mason simulation.}
    \label{fig:histogram}
\end{figure}

Subsequently, for each query produced by Mason, our SeqAn program generates all possible 10-mer minimizers in the query and looks up their locations in the hash-table index of reference 10-mers. Thus, for each minimizer of the query, we found every occurrence of the minimizer recorded in the index. For each of these occurrences, we then generated a candidate alignment by selecting a region of the reference around the location indicated in the index. We selected a region 15\% larger than the query, and used the location of the minimizer in the query to determine how much of this region would occur before the location in the index and how much would occur after. A distribution of the number of base pairs in the candidates corresponding to each 300-base pair query is shown in Figure \ref{fig:histogram}. Ultimately, we evaluated the performance of our edit distance acceleration using this dataset of 500 queries of length 300 base pairs, each with a variable number of associated candidates capped at 32K candidates per query. 


To evaluate performance, we compared the end-to-end execution time for performing the Myers' algorithm on the APU with an optimized CPU implementation. We ran the CPU version on an Intel Xeon Gold 6230R CPU @ 2.1GHz with DDR4-2933. All results were averaged over ten runs. 

%% file: sec-results.tex

\section{Results}
\label{sec-results}

Using the dataset described above, we first explored APU performance relative to single-core CPU performance. The APU provides a total average speedup of 14.1$\times$ relative to single-core CPU performance on the Intel Xeon Gold 6230R processor, with a standard deviation of 0.07. 

\subsection{Execution Time Breakdown}

\begin{table}
  {\centering
  \cbxsetfontsize{9pt}

  \caption{Total Execution Time Breakdown}
  \label{tab:execution-breakdown}

  \setlength{\tabcolsep}{3pt}
  \begin{tabular}{l
    >{\raggedleft\arraybackslash}p{0.85in}
    >{\raggedleft\arraybackslash}p{0.85in}}
  \toprule
    \normalfont\BF{Function}         & \BF{Single-Core CPU (ms)}    & \BF{Single-Core APU (ms)}    \\                                     \toprule           
    initialize read params         &  <0.01       & <0.01 \\
    allocate on shared DRAM        &   0.00       &  0.01  \\
    initialize parameters          &  <0.01       & <0.01 \\
    initialize peq array           &  <0.01       &  0.36 \\
    copy data host to device       &   0.00       &  0.60 \\
    kernel                         & 152.73       &  9.67 \\
    copy data device to host       &   0.00       &  0.19 \\
    free allocated memory          &   0.00       &  0.01 \\
    \midrule
    total                          & 152.73       & 10.82 \\
  \bottomrule
  \end{tabular}\par}

  \vspace{0.05in}
\end{table}


Table \ref{tab:execution-breakdown} shows a breakdown of the time spent in each portion of the end-to-end execution time for filtering. As seen in the table, for both the CPU and APU, the actual time spent performing computations on the device (labeled "kernel time") dominates the end-to-end execution time of the program. The kernel alone gains a speedup up of 15.8$\times$ on the accelerator relative to the CPU. 

\begin{table}
  {\centering
  \cbxsetfontsize{9pt}

  \caption{Kernel Time Breakdown}
  \label{tab:kernel-breakdown}

  \setlength{\tabcolsep}{3pt}
  \begin{tabular}{l
    >{\raggedleft\arraybackslash}p{0.85in}
    >{\raggedleft\arraybackslash}p{0.85in}}
  \toprule
    \normalfont\BF{Code Section}         & \BF{Avg \# of Cycles (1K)*}     & \BF{Percent of Total Cycles}    \\                                     \toprule           
    loading saved Pv and Mv        &  856.2    &   8.7 \\
    computing eq                   & 1329.0    &  13.6 \\
    computing Xv                   &  605.1    &   6.2 \\
    computing Xh                   & 1119.6    &  11.4 \\
    computing Ph                   &  769.3    &   7.9 \\
    computing Mh                   &  512.5    &   5.2 \\
    computing scores               &  432.0    &   4.4 \\
    shift and save Ph              &  955.5    &   9.8 \\
    shift and save Mh              &  787.5    &   8.0 \\
    computing Pv                   &  735.0    &   7.5 \\
    computing Mv                   &  872.6    &   8.9 \\
    storing Pv and Mv              &  813.9    &   8.3 \\
  \bottomrule
  \end{tabular}\par}

  \vspace{0.05in}

  \cbxsetfontsize{9pt}

  *~Indicates total cumulative cycle count (in thousands) for all times each section of code was executed. All numbers refer to computing Myers' on one single query/candidate set, averaged over 10 trials. 
  
\end{table}


To provide deeper insights, Table \ref{tab:kernel-breakdown} shows a breakdown of the number of cycles spent in each portion of the computation kernel when running on the APU. The code sections closely correspond to the algorithm pseudocode in Figure \ref{fig:microcode}(a). As seen in the table, the largest number of cycles is devoted to computing the \(eq\) vectors. This is done at the beginning of each iteration of the innermost loop of the algorithm, and requires multiple sets of vector-search operations followed by masking operations. Computing this on the device is much faster than doing it on the CPU before launching the kernel, but still contributes a large number of cycles to the total performance. The second-largest cycle count belongs to the computation of the \(Xh\) vectors. The \(Xh\) vector is an intermediate result that propagates results horizontally along rows of the traditional score matrix, where each element depends on the one before it. In the accelerator's data layout, this means each bit of each element of the \(Xh\) vector depends on the bit before it. In order to calculate \(Xh\) efficiently, we follow \cite{myers-algorithm-1999} in using an addition operation to leverage its bit-carry property. There may be opportunities to further optimize this that future work can explore. 


The total end-to-end speedup is slightly lower than that of just the kernel due to other overheads, including the overhead of moving data to and from the host and device and some limited data initialization on the host. As shown in the table, the second-largest portion of the accelerator execution time after the kernel is the time spent copying data from the host to the device. The bit-packed representation of the candidate sequence data described in Section \ref{sec-acceleration} has reduced this to the value seen here. The next-largest contributor of overhead is the time spent computing the \(peq\) matrix on the host processor before launching the kernel to the device. Since the data would have to be transposed to be useful in the main kernel of computation, it was pre-computed on the host processor. Future work can explore efficient ways to migrate this computation to the device. 

\subsection{Parameter Sweeping}

\begin{figure}
    \centering
    \includegraphics[width=\columnwidth]{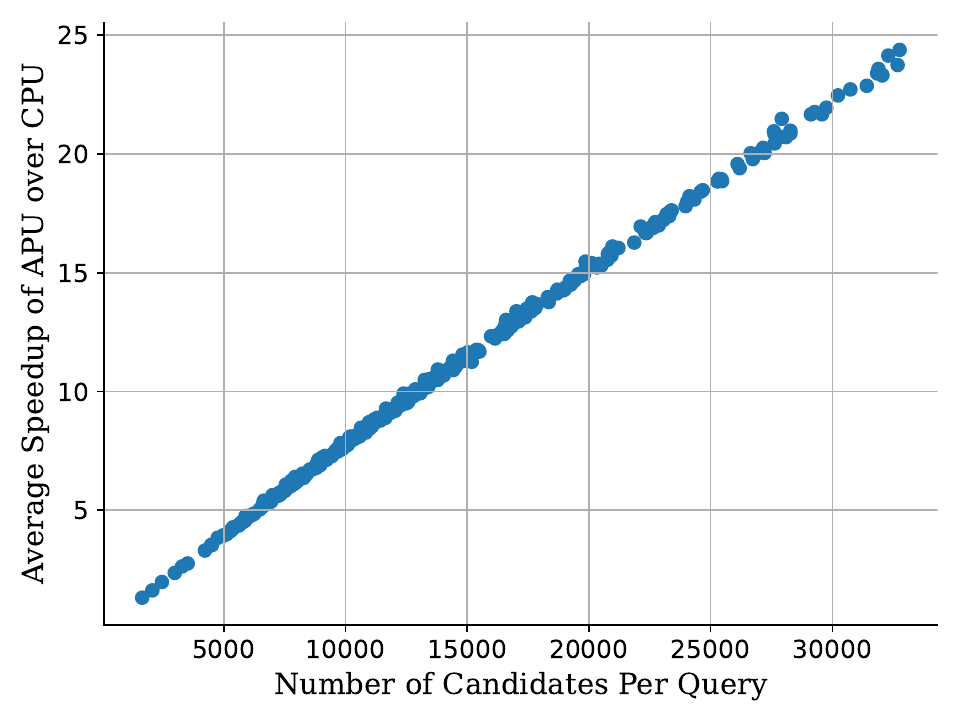}
    \caption{Speedup of APU over CPU for Each Query}
    \label{fig:seed-sweep}
\end{figure}

Since candidate alignments are done in parallel for a given query, the greatest performance benefit of the APU is realized when all 32K columns of the SRAM are utilized. This corresponds to aligning a single query against 32K candidate sequences. However, as seen in the histogram in Figure \ref{fig:histogram}, each query produces a variable number of candidates that is often less than 32K. To understand more deeply how the relative performance of the APU and CPU change with the number of candidates, we plotted the average speedup for each individual query as a function of its number of associated candidates. The results can be seen in Figure \ref{fig:seed-sweep}.

When the number of candidates increases, the execution time of the kernel (actual computations) on the APU stays almost constant, as operations are done in parallel across all 32K columns of the SRAM regardless. There is a slight linear increase in total APU execution time due to copying more data from the host to device and back. However, the CPU execution time increases linearly much faster with the number of candidates, as the processor must loop over every candidate for each query. Since the CPU execution time increases faster than the APU execution time, speedup increases. This trend can be seen in the graph.

\begin{figure}
    \centering
    \includegraphics[width=\columnwidth]{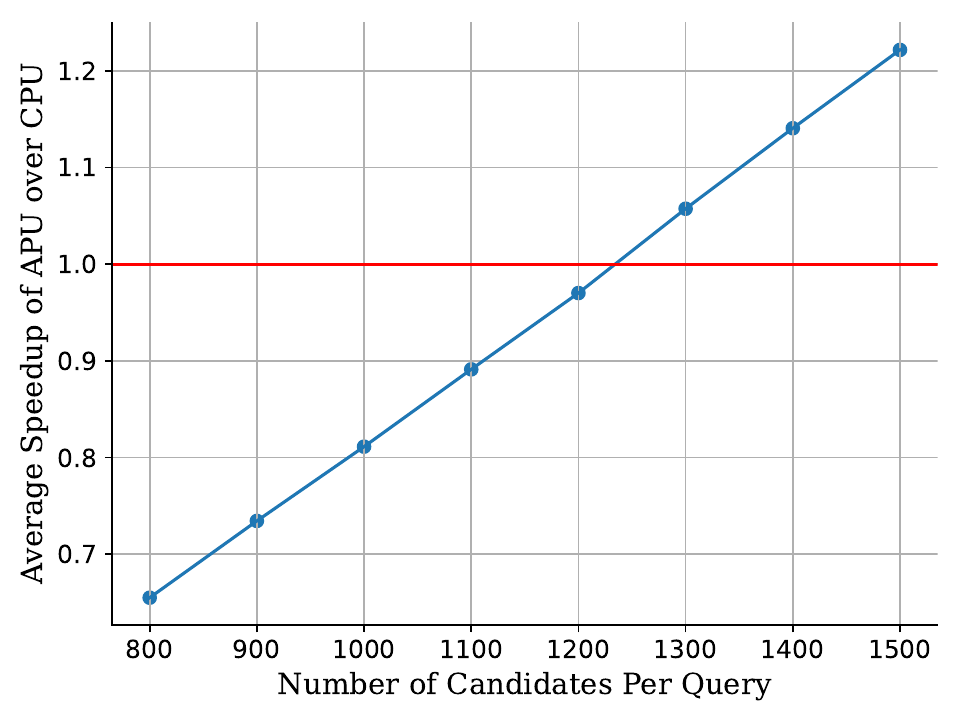}
    \caption{Fine-Grained Sweep of APU Speedup}
    \label{fig:seed-sweep-synthetic}
\end{figure}

At what number of candidates per query does the APU become more optimal than the CPU? To answer this, we ran a set of experiments that capped the number of candidates per query at various values between 800 and 1800. The results are shown in Figure \ref{fig:seed-sweep-synthetic}. Having at least 1300 candidates per query allows the APU to be faster than the CPU. Thus, even if earlier steps in the read-assembly pipeline were more optimized, using techniques such as chaining during candidate generation to produce only more likely alignment locations, the APU would likely still provide speedup over a single-core CPU. 

For the query and candidate lengths used here, at its optimal performance (i.e., when the number of candidates for a single query is 32K), the APU provides average speedups of 24.1$\times$, with a standard deviation of 1.28.

As demonstrated above, the APU has the capacity to filter a large number of candidates per query sequence. This could relieve pressure on earlier filtering or chaining steps in the pipeline, since extra parallelism here incurs very little overhead. 

\subsection{Multicore Results}

The CPU baseline can also be multithreaded to take advantage of all 16 cores of the Intel CPU chip. We parallelized our CPU baseline over different queries, so each core processes a chunk of queries and aligns all associated candidates against each query. Using 16 cores in this manner results in an average speedup of 1.49$\times$ for the APU over the CPU. 

However, there are also four cores available on the APU, of which the current edit distance implementation only uses one. We expect a roughly linear multicore speedup on the APU (similar to what was seen with the CPU), leading to an estimated total speedup of 5-6$\times$ for a 4-core APU over a 16-core CPU. 

%% file: sec-related.tex

\section{Related Work}
\label{sec-related}

There have been numerous approaches to accelerating reference-guided DNA assembly. Most prior work focuses on using hardware acceleration for the final gap-affine alignment step because it is the most computationally expensive. Such work includes custom hardware accelerators such as Darwin~\cite{turakhia-darwin-asplos2018} and GenAx~\cite{fujiki-genax-isca2018}. SeedEx further reduces alignment time by adopting a heuristic algorithm called banded Smith-Waterman ~\cite{fujiki-seedex-micro2020}. SeGraM extends the acceleration of sequence-to-sequence tasks into sequence-to-graph tasks, with acceleration still focused on the gap-affine alignment phase~\cite{cali-segram-isca2022}. Some recent work has begun to use compute-in-memory for acceleration of this gap-affine alignment step, such as GenCache~\cite{nag-gencache-micro2019}. GenCache integrates in-cache operators into the GenAx accelerator to accelerate a custom Silla gap-affine alignment algorithm. 

An alternative approach accelerates edit distance and uses it for that same final sequence alignment step. This has higher performance but may have lower accuracy than gap-affine scoring, making it optimal for some biological applications but not applicable to more complex analysis of proteins and structural variation. Such work includes the GenASM accelerator~\cite{cali-genasm-micro2020} that proposed hardware for the Bitap edit distance algorithm, a different bit-vector algorithm for sequence alignment. 

Recent work has begun to explore accelerating the filtering step of short-read DNA assembly, as we have done in this paper, in order to reduce the number of alignments that need to be performed in the more expensive gap-affine alignment stage. Shouji~\cite{alser-shouji-bioinformatics2019} and GateKeeper~\cite{alser-gatekeeper-bioinformatics2017} use FPGA-based accelerators. GRIM-Filter, the most closely related work, uses a simulation-based study of 3D-stacked compute-in-DRAM to accelerate a custom filtering algorithm based on kmer-matching~\cite{kim-grimfilter-2018}. While both Myers' and the GRIM-Filter algorithm are effective in accurately filtering query/candidate pairs, the Myers' algorithm's massively parallel bit-operations make it more suitable for the APU architecture described here. 

As a promising accelerator for genomics workloads, the APU also represents the first, to our knowledge, commercial compute-in-SRAM system. Compute-in-SRAM has been previously only explored in simulation or small-scale academic prototypes. Bitline computing was introduced in SRAM by Jeloka et al.~using 28nm academic prototypes~\cite{jeloka-compute-in-sram-vlsic2015, jeloka-compute-in-sram-jssc2016}. A related line of work built on this to provide the ability to perform floating-point operations for neural networks~\cite{eckert-neural-caches-isca2018}, performing computation in horizontal bitlines~\cite{wang-compute-in-sram-isscc2019}, and adopting a SIMT abstraction to accelerate workloads in a data-parallel manner~\cite{fujiki-duality-caches-isca2019}. Bit-serial and bit-parallel operations were further explored in VRAM~\cite{alhawaj-eve-circuits-iscas2020}. 

%% file: sec-conclusion.tex

\section{Conclusion}
\label{sec-conclusion}

In this paper, we illustrated promising on-going work that accelerates the filtering step of reference-guided DNA assembly using commercial compute-in-SRAM. We demonstrated the ability of the Gemini APU to provide average speedups for the Myers' bit-parallel edit distance algorithm of 14.1$\times$ relative to single-core Intel Xeon Gold 6230R CPU performance. 

Several key characteristics emerge as features of the APU that make the system well-suited to Myers' algorithm and likely to other genomics workloads as well. First, as demonstrated by the increasing speedups that result from increasing seed count, the massive parallelism of the APU is key. Secondly, bit-parallel operations like those found in Myers' algorithm allow the application to take full advantage of this parallelism. Bitwise operations can be performed with high computational throughput using the low-level microcode primitives we described. In addition, many genomics applications use low-precision integer arithmetic operations, such as on DNA datasets that can be encoded as two-bit values. This allows the programmer to pack more data into 16-bit integers, as we did here. Finally, this particular application has a high degree of data reuse, but other applications will need to carefully consider the ratio of data movement to computation. Future work can explore more complex genomics algorithms that exploit these properties. 

The APU is thus a promising way to accelerate filtering and likely other genomics workloads. As in the case of filtering, the massive parallelism the APU provides may prompt rethinking of traditional genomics algorithms or their context within the genomics pipeline. 